\documentclass[twocolumn,pre,superscriptaddress]{revtex4}

\usepackage{graphicx}
\usepackage{amsmath,amssymb}
\usepackage{color}
\usepackage{epstopdf}

\newcommand\defn{\textit}

\newcommand\set[1]{\lbrace#1\rbrace}

\begin{document}

\title{Large-scale structure of time evolving citation networks}

\author{E. A. Leicht}
\affiliation{Department of Physics, University of Michigan, Ann Arbor,
MI 48109, U.S.A.}
\author{Gavin Clarkson}
\affiliation{School of Information, University of Michigan, Ann Arbor,
MI 48109, U.S.A.}
\author{Kerby Shedden}
\affiliation{Department of Statistics, University of Michigan, Ann Arbor,
MI 48109, U.S.A.}
\author{M. E. J. Newman}
\affiliation{Department of Physics, University of Michigan, Ann Arbor,
MI 48109, U.S.A.}
\affiliation{Center for the Study of Complex Systems, University of
  Michigan, Ann Arbor, MI 48109, U.S.A.}

\begin{abstract}
  In this paper we examine a number of methods for probing and
  understanding the large-scale structure of networks that evolve over
  time.  We focus in particular on citation networks, networks of
  references between documents such as papers, patents, or court cases.  We
  describe three different methods of analysis, one based on an
  expectation-maximization algorithm, one based on modularity optimization,
  and one based on eigenvector centrality.  Using the network of citations
  between opinions of the United States Supreme Court as an example, we
  demonstrate how each of these methods can reveal significant structural
  divisions in the network, and how, ultimately, the combination of all
  three can help us develop a coherent overall picture of the network's
  shape.
\end{abstract}

%\pacs{PUT PACS HERE}

\maketitle

\section{Introduction}
\label{sec:introduction}

The physics community has in recent years devoted considerable attention to
the study of networks, including social networks, biological networks,
information networks, and others~\cite{AB2002,DM2002,NewmanSIAM}.  Many of
these networks also have long histories of study in other fields.  Citation
networks, which are the principal focus of this paper, have been studied
quantitatively almost from the moment citation databases first became
available, perhaps most famously by the physicist-turned-science-historian
Derek de Solla Price, who authored two celebrated papers in the 1960s and
1970s highlighting the power-law degree distributions in networks of
scientific papers and developing models to explain their
origin~\cite{Price1965,Price1976}.

A citation network is an information network in which the vertices
represent documents of some kind and the edges between them represent
citation of one document by another.  Citation networks differ from other
networks in a number of important ways.  First, they are directed:
citations go from one document to another and hence constitute an
inherently asymmetric relationship between the vertices involved.
Mathematically, the network can be represented by an adjacency
matrix~$\mathbf{A}$, with elements
\begin{equation}
A_{ij} = \left\lbrace\begin{array}{ll}
           1 & \qquad\mbox{if there is an edge from $j$ to $i$,}\\
           0 & \qquad\mbox{otherwise.}
         \end{array}\right.
\end{equation}
In a directed network the adjacency matrix is, in general, asymmetric.

\begin{figure}
\includegraphics[width=4.5cm]{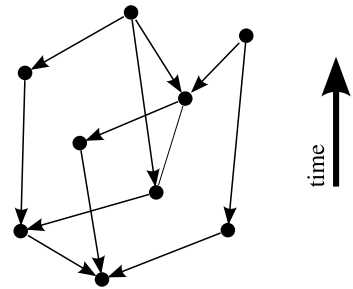}
\caption{Citations run from vertices created at later times to those
  created at earlier times---in the opposite direction to ``arrow of
  time.''}
\label{fig:citationNetworkCartoon}
\end{figure}

A second feature of citation networks is that they evolve over time as new
documents are created.  The time evolution of the network takes a special
form, in that vertices and edges are added to the network at a specific
time and cannot be removed later.  This permanence of vertices and edges
means that the structure of the network is mostly static: it changes only
at the ``leading edge'' of the network, the current time at which new
documents are being added.  Citation networks differ in this respect from
other information networks such as the world wide web, in which vertices
and edges can be removed as well as added and edges can be repositioned
after they are added.  The limited form of time evolution found in citation
networks makes them, in some ways, a simpler and cleaner laboratory for the
study of network growth than the web.

The combination of the two features of citation networks described above
leads to a third: citation networks are acyclic, meaning there are no
closed loops of citations of the form A cites B cites C cites A, or longer.
When a new vertex is added to a citation network it can cite any of the
previously existing vertices, but it cannot cite vertices that have not yet
been created.  This gives the network a clear ``arrow of time,'' with all
edges pointing backwards in time as shown in
Fig.~\ref{fig:citationNetworkCartoon}.  As a result it is typically
possible, starting from a given vertex, to find a path of citations that
takes us back in time through the network, but it is not possible to find
one that takes us forward again, so that no closed loops exist.  (Real
citation networks are often not perfectly acyclic.  For example, a
scientific paper can sometimes cite work that is forthcoming but not yet
published, resulting in a closed loop in the network.  However, such loops
are rare and necessarily short, being limited by the narrow span of time
over which it is possible to predict future publications.  In practice,
therefore, it is usually a good approximation to assume the network to be
acyclic.)

Citation networks arise in a variety of different areas.  We have mentioned
networks of scientific citations, which have been studied by many authors
since the classic work of Price mentioned above.  (See, for instance, the
book by Egghe and Rousseau~\cite{ER90} or any volume of the journal
\textit{Scientometrics}, which is entirely devoted to the quantitative
analysis of scholarly authorship and citation patterns.)  Citation networks
of patents have, to a lesser extent, also been studied.  Patents cite other
patents for a variety of reasons, but most often to establish their
originality and distinction from previous work.  Extensive data on patent
citations have become available in recent years, allowing the construction
of very large citation networks~\cite{JT02,CD06}.  Very recently, there has
also been interest in legal citation networks, networks of legal opinions
written by judges and others, which cite one another to establish
precedent~\cite{CrossAug2006,Fowler07}.  We make extensive use of one
particular legal citation network, the network of opinions of the United
States Supreme Court, as an example in this paper, although the techniques
we will be considering are certainly applicable to other networks as well.

Given the wide interest in and unique structure of citation networks, it is
instructive to investigate what can be learned from an analysis of the
statistical patterns present in these networks.  A variety of studies have
been presented in the past focusing on relatively standard network measures
such as degree distributions~\cite{Price65,Seglen92,Redner98}.  To
investigate the time-dependent structure that is the special property of
citation networks, however, other methods are needed.  In this paper we
present several techniques that, as we will show, are---both individually
and collectively---capable of revealing interesting new structure in these
networks.

\section{A mixture model of citation patterns}
\label{sec:em}
The first analysis we describe makes use of a stochastic mixture model of
the citation process, which is fitted to the observed network data using
the likelihood optimization technique known as the expectation-maximization
algorithm.

A crucial property affecting the structure of citation networks is the
pattern over time of the citation of documents following their publication.
It is interesting for instance to ask if there are typical patterns that
documents follow.  Are there more citations immediately after publication
than later, or do they grow in frequency over time?  Are documents more
likely to cite recent precedents or older better-established ones?  Do
documents tend to cite others published during a particular time period?
There could also be more than one common pattern with different documents
following different patterns.  If so, how can we determine those patterns,
and how can we tell which pattern particular documents follow, given that
citation data are inherently noisy?

As an example, we consider the network of legal citations between cases
handed down by the Supreme Court of the United States, from its inception
in 1789 until the present day.  We will use this example throughout this
paper; it is well documented, shows clear and interesting structural
signatures, and has been studied much less than other types of citation
networks in the past, so that, although we use the network primarily as an
illustrative device, the results we derive are in many cases of interest in
their own right and not just as a demonstration of our methods.

Consider the following table, which gives the dates of the citations
received so far by a single example opinion handed down by the Supreme
Court in the year 1900:
\begin{center}
\setlength{\tabcolsep}{6pt}
\begin{tabular}{rr|rr|rr}
  year & cites & year & cites & year & cites \\
\hline
  1900 & 1 & 1907 & 2 & 1925 & 1\\
  1901 & 4 & 1910 & 1 & 1936 & 1\\
  1902 & 3 & 1912 & 2 & 1947 & 1 \\
  1904 & 1 & 1920 & 1\\
\end{tabular}
\end{center}
We will take citation profiles such as this as the basic inputs in
our analysis.

One interesting question (there are many) is whether there are distinct
eras of citation in the history of this (or any) citation network.  Are
there, for instance, eras in which a certain set of documents are well
cited, followed perhaps by another era or eras in which that set falls out
of favor to be replaced by a different one?  Many readers can probably
think of anecdotal cases of behavior like this in scientific citation
networks.  Here we place these observations on a firm analytic foundation.

We will attempt to divide the vertices in a citation network into groups by
identifying similarities in their citation profiles.  Our method will be to
define a set of citation profiles and then self-consistently assign each
case to the profile it best fits while at the same time adjusting the shape
of the profiles to best fit the cases assigned to them.  The means by which
we accomplish this task is the expectation--maximization (EM)
algorithm~\cite{emPaper,emBook}.

The EM algorithm is an established tool of statistics, but one that is
relatively new to network analysis.  In a previous paper we described an
application of the method to the classification of vertices in static
networks, both directed and undirected~\cite{NL07}.  Here we describe a
different application to the analysis of the temporal profiles of
citations.

In essence the EM algorithm is a method for fitting a model to observed
data by likelihood maximization, but differs from the maximum likelihood
methods most often encountered in the physics literature in that it does
not rely upon Markov chain Monte Carlo sampling of model parameters.
Instead, by judicious use of ``hidden'' variables, the maximization is
performed analytically, resulting in a self-consistent solution for the
best-fit parameters that can be evaluated using a relatively simple
iteration scheme.

Suppose we have a network of $n$ vertices representing our documents and we
believe that they can be divided into $c$ groups, each of which is
characterized by a particular probability distribution of citations over
time.  (Ultimately, we will vary $c$ to find the best description for our
data, but for the moment let us assume it to be fixed.)  Our approach to
finding the groups will be to fit the network to a model consisting of two
parts: (1)~a set of \defn{time profiles}~$\set{\theta_r(t)}$, one for each
group, such that $\theta_r(t)$ is the probability that a particular
citation received by a document in group~$r$ is made during year~$t$; (2)~a
set of \defn{probabilities}~$\pi_r$, such that $\pi_r$ is the probability
that a randomly chosen document belongs to group~$r$ (i.e.,~$\pi_r$ is the
expected fraction of documents belonging to group~$r$).  We fit this model
to the observed data by maximizing the probability of the observed set of
citations given the model---the so-called likelihood function.

Suppose that document~$i$ belongs to group~$g_i$ and let $z_i(t)$ be the
number of citations that the document receives in year~$t$.  Then the
probability that document~$i$ received the particular citations it did and
is in group~$g_i$, given the model parameters, is
\begin{equation}
\Pr(z_i,g_i|\pi,\theta) = \Pr(z_i|g_i,\pi,\theta) \Pr(g_i|\pi,\theta),
\label{eq:prpr}
\end{equation}
where for convenience we use $\pi,\theta$ to denote the entire set
$\set{\pi_r,\theta_r}$.  Assuming random and uncorrelated citations drawn
from the time profile~$\theta_{g_i}(t)$, the terms on the right-hand side
are given by
\begin{align}
\label{eq:pztheta}
\Pr(z_i|g_i,\pi,\theta) &= k_i! \prod_{t=t_1}^{t_2} {\left[ \theta_{g_i}(t)
    \right]^{z_i(t)} \over z_i(t)!},\\
\label{eq:pgi}
\Pr(g_i|\pi,\theta) &= \pi_{g_i},
\end{align}
where $k_i=\sum_t z_i(t)$ is the in-degree of document~$i$, i.e.,~the total
number of citations it receives, and $t_1$~and $t_2$ are the first and last
years of data in our dataset.

Now taking the product over all vertices, the likelihood of the entire data
set is $L = \prod_{i=1}^n \Pr(z_i,g_i|\pi,\theta)$.  In fact, we will work
with the logarithm~$\mathcal{L}$ of the likelihood, which has its maximum
in the same place:
\begin{equation}
\mathcal{L} = \ln L = \sum_{i=1}^n \bigl[ \ln \Pr(g_i|\pi,\theta)
                      + \ln \Pr(z_i|g_i,\pi,\theta) \bigl].
\label{eq:ll1}
\end{equation}
Unfortunately, $\mathcal{L}$~depends on the group memberships~$g_i$, which
we don't know.  Given the observed citation patterns, however, we can make
a good guess about the group memberships, or more precisely we can compute
the probability distribution of their values, which in Bayesian fashion we
regard as a statement about our knowledge of the world, rather than a
statement about the actual values of the group memberships, which are in
theory perfectly well-defined quantities.  Writing the probability of a
particular assignment of vertices to groups as
$\Pr(\set{g_i}|z,\pi,\theta)$, we can then calculate the expected value of
the log-likelihood as the average of Eq.~\eqref{eq:ll1} over all possible
assignments thus:
\begin{align}
\overline{\mathcal{L}} &=
  \sum_{g_1=1}^c\ldots\sum_{g_n=1}^c \Pr(\set{g_i}|z,\pi,\theta)\,
  \mathcal{L} \nonumber\\
  &= \sum_{g_1=1}^c\ldots\sum_{g_n=1}^c \Pr(\set{g_i}|z,\pi,\theta)\nonumber\\
  &\qquad{}\times \sum_{i=1}^n \bigl[ \ln \Pr(g_i|\pi,\theta) + 
                  \ln \Pr(z_i|g_i,\pi,\theta) \bigr] \nonumber\\
  &= \sum_{i=1}^n \sum_{r=1}^c \Pr(g_i=r|z_i,\pi,\theta) \nonumber\\
  &\qquad{}\times \bigl[ \ln \Pr(g_i=r|\pi,\theta) + 
                  \ln \Pr(z_i|g_i=r,\pi,\theta) \bigr] \nonumber\\
  &= \sum_{i=1}^n \sum_{r=1}^c q_{ir}
     \Bigl\lbrace \ln \pi_r + \ln k_i! + {} \nonumber\\
  &\hspace{6em} \sum_{t=t_1}^{t_2} \bigl[ z_i(t) \ln \theta_r(t) -
                                    \ln z_i(t)! \bigr] \Bigl\rbrace,
\label{eq:loglikelihood}
\end{align}
where we have introduced the shorthand notation
\begin{equation}
q_{ir} = \Pr(g_i=r|z_i,\pi,\theta)
\end{equation}
for the probability that vertex~$i$ belongs to group~$r$, given the model
and the observed citation pattern.

This expected log-likelihood represents our best estimate of the value of
the log-likelihood given what we know about the system.  By maximizing it,
we can now calculate a best estimate of the most likely values of the model
parameters, a process that involves two steps: first, we estimate the group
membership probabilities~$q_{ir}$; second, we use those probabilities in
the maximization of~$\overline{\mathcal{L}}$.  We take these steps in turn.

To calculate the $q_{ir}$ we observe that
\begin{equation}
q_{ir} = \Pr(g_i=r|z_i,\pi,\theta)
       = {\Pr(z_i,g_i=r|\pi,\theta)\over\Pr(z_i|\pi,\theta)}.
\end{equation}
The two factors on the right can be determined by summing
Eq.~\eqref{eq:prpr} over the appropriate sets of variables and making use
of Eqs.~\eqref{eq:pztheta} and~\eqref{eq:pgi} to give
\begin{equation}
q_{ir} = {\pi_r \prod_t \left[ \theta_r(t) \right]^{z_i(t)}\over
         \sum_k \pi_k \prod_t \left[ \theta_k(t) \right]^{z_i(t)}}.
\label{eq:estep}
\end{equation}

Once we have this expression, we can use it to evaluate the log-likelihood,
Eq.~\eqref{eq:loglikelihood}, and hence to find the values of the model
parameters that maximize the likelihood, which is our ultimate goal.  The
maximization is helped by the fact that $\pi_r$ and $\theta_r$ enter
Eq.~\eqref{eq:loglikelihood} in independent terms.  Considering $\pi_r$
first and noting that it must satisfy the normalization condition~$\sum_r
\pi_r=1$, we introduce a Lagrange multiplier~$\alpha$ and then
differentiate, holding $q_{ir}$ constant, to get
\begin{eqnarray}
0 &=& {\partial\over\partial\pi_r} \biggl\lbrace \sum_{ir} q_{ir}
      \ln\pi_r + \alpha \biggl[ 1 - \sum_r \pi_r \biggr] \biggr\rbrace
      \nonumber\\
  &=& {1\over\pi_r} \sum_{i=1}^n q_{ir} - \alpha.
\end{eqnarray}
Rearranging this expression gives
\begin{equation}
\pi_r = {1\over\alpha} \sum_{i=1}^n q_{ir}.
\end{equation}
The Lagrange multiplier~$\alpha$ is then fixed by the condition $\sum_r
\pi_r=1$ thus:
\begin{equation}
\sum_{r=1}^c \pi_r = 1 = {1\over\alpha} \sum_{ir} q_{ir} = {n\over\alpha},
\end{equation}
where we have made use of $\sum_r q_{ir} = 1$.  Thus $\pi_r$ is given by
\begin{equation}
  \pi_r = {1\over n} \sum_i q_{ir}.
\label{eq:mstep1}
\end{equation}
In other words, the prior probability of a vertex belonging to group~$r$ is
just the average over all vertices of the conditional probability of
belonging to group~$r$.

Similarly, the~$\theta_r$ satisfy the normalization condition $\sum_t
\theta_r(t)=1$ for all~$r$, so we introduce a set of $c$ Lagrange
multipliers~$\set{\beta_r}$ and write
\begin{eqnarray}
  & & {\partial\over\partial\theta_r(t)} \biggl\lbrace \sum_{ir}
      q_{ir} \sum_{t=t_1}^{t_2} z_i(t) \ln \theta_r(t) \nonumber\\
  & & {} \hspace{6em} + \sum_r \beta_r \biggl[ 1 - \sum_t \theta_r(t) \biggr]
      \biggr\rbrace = 0.
\end{eqnarray}
Again holding $q_{ir}$ constant and employing Eq.~\eqref{eq:pztheta}, we
find
\begin{equation}
\sum_i q_{ir} {z_i(t)\over\theta_r(t)} - \beta_r = 0,
\end{equation}
or
\begin{equation}
\theta_r(t) = {\sum_i q_{ir} z_i(t)\over\sum_i q_{ir} k_i},
\label{eq:mstep2}
\end{equation}
where we have evaluated $\beta_r$ using the normalization condition and the
fact that $\sum_t z_i(t) = k_i$ by definition.

To calculate the optimal values of the model parameters, as well as the
group membership variables~$q_{ir}$, we now need to solve
Eq.~\eqref{eq:estep} simultaneously with Eqs.~\eqref{eq:mstep1}
and~\eqref{eq:mstep2}.  The simplest way to do this is numerical iteration.
Starting from an initial guess about the values
of~$\set{\pi_r,\theta_r(t)}$, we evaluate Eq.~\eqref{eq:estep} and then use
the results to make an improved estimate of the model parameters from
Eqs.~\eqref{eq:mstep1} and~\eqref{eq:mstep2}.  Under reasonable conditions
this process is known to converge upon iteration to a self-consistent
solution.

\subsection{Example}
\begin{figure}
\includegraphics[width=8cm]{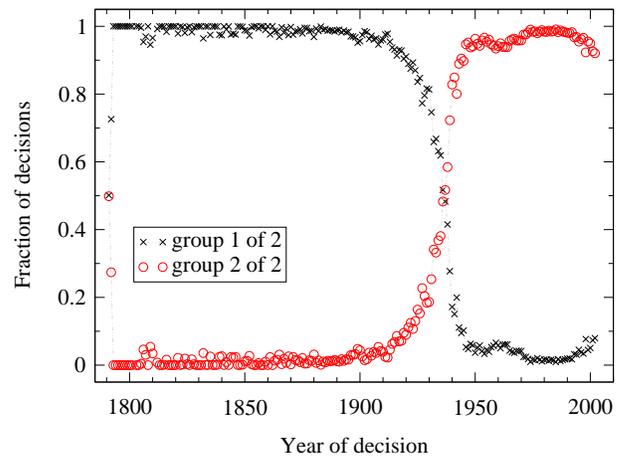}
\caption{Results of the application of the EM analysis with $c=2$ to the
  network of citations between Supreme Court opinions.  The two curves show
  the fraction of cases assigned to each of the two groups found, as a
  function of time.}
\label{fig:emsplit2}
\end{figure}

As a demonstration of the EM method we have applied it to the citation
network of Supreme Court cases described in Section~\ref{sec:em}.  Applied
to this network, the algorithm will divide the network into any requested
number~$c$ of groups, such that each group is characterized by a
distinctive pattern of citations to cases in that group.  We have performed
the analysis for a variety of different values of~$c$.  We begin with the
simplest case, $c=2$, of division into two groups.  Starting with random
initial values for $\set{\pi_r,\theta_r}$ and applying the EM iteration,
Eqs.~\eqref{eq:estep}, \eqref{eq:mstep1}, and~\eqref{eq:mstep2}, the
parameters rapidly converge to a clear split of the network into two
groups.  Figure~\ref{fig:emsplit2} shows the fraction of cases assigned by
the algorithm to each of the groups as a function of time.  Cases are
assigned in proportion to their probability of membership in each of the
groups so that, for instance, a case belonging to group~1 with probability
0.7 and to group~2 with probability 0.3 contributes 0.7 of a case to the
first group and 0.3 of a case to the second.

Figure~\ref{fig:emsplit2} reveals a dramatic split between the two groups:
the best fit, in the maximum likelihood sense, of the mixture model with
two groups to these data produces one group containing practically all
cases before 1937 and another containing practically all cases after.  This
breakpoint coincides with a significant constitutional crisis for the
Supreme Court.  For the interested reader we give some further analysis in
Section~\ref{sec:discussion}.

\begin{figure}
\includegraphics[width=8cm]{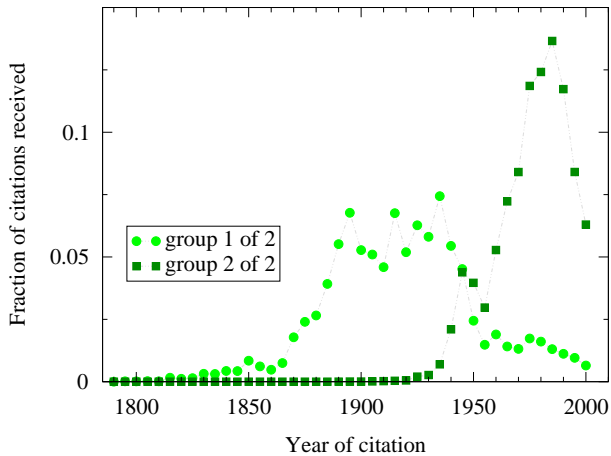}
\caption{The citation profiles~$\theta_r(t)$ generated by the EM algorithm
  with $c=2$ for the Supreme Court citation network.}
  \label{fig:emprofiles}
\end{figure}

The EM algorithm tells us in this case that the Supreme Court's rulings
split quite cleanly into groups with distinct citation profiles.  That is,
the opinions of the court can be distinguished sharply by the cases that
later cited them.  The citation profiles themselves, meaning the temporal
citation patterns represented by the parameters~$\set{\theta_r}$ in the
model, are shown in Fig.~\ref{fig:emprofiles}.  As we can see, they also
divide into two time periods, which correspond closely to those of the
group memberships depicted in Fig.~\ref{fig:emsplit2}.  This implies that
the opinions that cite cases in each of our groups were handed down during
roughly the same eras as the cited cases.  This is not surprising if one
assumes that the group divisions reflect different legal ideologies, but it
is important to bear in mind that our analysis does not require it: it
would be perfectly possible to detect groups that were distinguished by
citations received during some entirely different era of the court
arbitrarily later in its history, or even in no era at all but scattered
widely over time.

We can also ask about best fits to the model for numbers of groups~$c$
greater than two.  It is always the case that larger values of~$c$ will
give better fits to the data, since larger values give us more parameters
to fit with, but we must be wary of overfitting.  In practice, we have been
able to extract useful formation about networks by comparing the results
for a variety of small values of~$c$.  Rigorous methods for deciding
optimal values of~$c$, such as minimum description length, methods based on
approximations to the marginal likelihood, or information theoretic
measures have been developed for other applications of the EM
algorithm~\cite{Akaike74,Schwarz78} and we discuss these approaches
elsewhere.  For the moment we simply describe the results for various
values of~$c$.

\begin{figure}
\includegraphics[width=8cm]{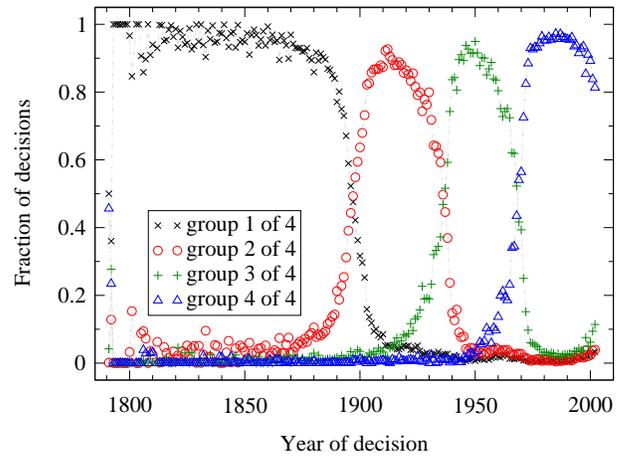}
\caption{Results of the application of the EM analysis with $c=4$ to the
  network of citations between Supreme Court opinions.}
\label{fig:emsplit3}
\end{figure}

Figure~\ref{fig:emsplit3} shows results for the Supreme Court network with
$c=4$.  The method again finds clear groups of cases, and as in the $c=2$
case they are strongly delineated according to the dates of the opinions
and thus appear to offer evidence for the presence of distinct eras in the
court's history.  In particular, the analysis finds a clear grouping of
cases between 1897 and 1937, corresponding approximately to the so-called
\textit{Lochner} era of Supreme Court jurisprudence, the significance of
which is described in Section~\ref{sec:discussion}.

\begin{figure}
\includegraphics[width=8cm]{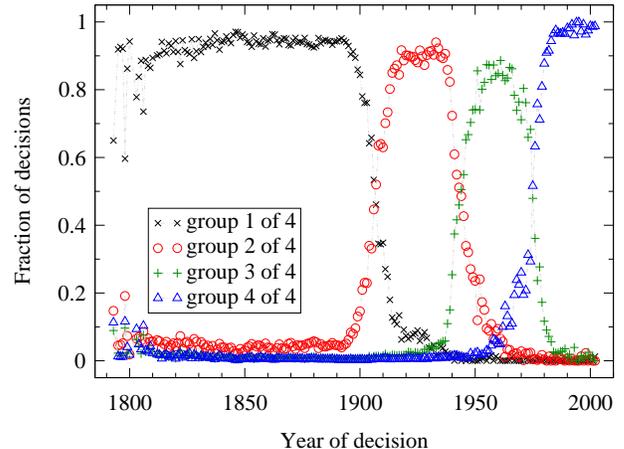}
\caption{Results of the application of the EM algorithm with $c=4$ to data
  for citations \emph{made} (rather than received) by opinions in our
  Supreme Court dataset.  The groups found are quite similar to those for
  the analysis based on citations received.}
\label{fig:emciting3}
\end{figure}

In these analyses we have characterized our documents by the pattern of
citations they receive.  However, one can equally well look at the pattern
of citations that documents \emph{make} and this also, at least in some
cases, can be a useful cue for detecting patterns in the network.  The EM
algorithm can be applied to this analysis as well.  The developments are
identical and the same computer code can be used---one simply takes the
transpose of the adjacency matrix.  Figure~\ref{fig:emciting3}, for
example, shows the results of the application of the method to citations
made by the opinions in our Supreme Court dataset, with $c=4$.  As the
figure shows, the results are remarkably similar to those for citations
received: it appears that, in this case at least, there is a high degree of
agreement about how cases should be classified into eras.  This could
indicate agreement between the opinions' writers and those that came after
them, about the position staked out by individual opinions within the
larger body of literature represented in our data set.

\section{Clustering in citation networks}
\label{sec:clustering}
The general problem of the division of networks into groups of related
vertices has been extensively studied in the past.  The classic problem of
``clustering'' or ``community detection'' is to find groups of vertices
within networks that have a higher than average density of internal edges
and relatively few connections to the rest of the
network~\cite{Newman04b,DDDA05}.  The second analysis technique we
investigate for citation networks is a clustering method of this kind.  As
we will see, it is instructive to compare the results with those of our EM
analysis in the previous section.  The two methods do not do the same
thing: the EM analysis groups together vertices that have similar time
profiles to their citations, while the community analysis groups together
vertices that are specifically linked to one another by edges.
Nonetheless, as we will show, the two approaches can produce similar
outcomes, for instance in the example of the Supreme Court data set.

Considerable effort has been devoted to the development of methods to find
community structure within networks.  The authors are aware of dozens of
different methods (at least) published within the last few years.  Here we
make use a method recently proposed by Newman~\cite{Newman06b} based on the
maximization of the benefit function known as ``modularity.''  Although
many competing methods appear to give excellent results, we focus on this
particular method for two reasons: first, it is based on firm statistical
principles that make its operation transparent to the user; second, it has
been shown in recent head-to-head comparisons to give better results on
standardized tests than competing methods~\cite{DDDA05}.

Briefly the method works as follows.  Given a network and a particular
division of the vertices of that network into nonoverlapping groups or
communities, the modularity is defined as the number of edges that lie
within those groups minus the expected number of such edges if edges are
placed at random between the vertices (but respecting vertex
degree)~\cite{NG04}.  In essence, the modularity measures whether a larger
than expected number of edges fall within the groups defined.  In
principle, the task of finding the best division of the network into groups
is then one of maximizing the modularity over all possible
divisions~\cite{Newman04a}.  In practice, this maximization problem is
known to be NP-complete~\cite{Brandes07}, so approximate solution methods
must be used for all but the smallest networks.  Newman's method works by
rewriting the modularity in the language of linear algebra as a quadratic
form involving an index vector and a characteristic matrix dubbed the
``modularity matrix.''  It can then be shown that the signs of the elements
of the leading eigenvector of this modularity matrix give an approximation
to the division of the network into two parts that maximizes the
modularity.  This approximate maximum can optionally be further refined by,
for instance, applying a greedy algorithm that moves vertices between
groups as described in~\cite{Newman06b}.  By repeatedly dividing the
network in two in this way, a network can be divided into any number of
communities, although typically one stops dividing when no divisions exist
that will increase the modularity any further.

This repeated subdivision of the network into smaller and smaller groups is
particularly attractive for the purposes of our present analysis, because
it allows us to observe the major divisions in the network first, followed
by more minor ones, and to stop the process at any point to compare with
our other analyses.  A limitation of the method is that it is designed for
use with undirected rather than directed networks.  This however is not a
great hindrance.  It seems reasonable to consider edges in a citation
network to be a sign of connection between documents, and that connection
exists regardless of the direction the edge runs in.  So we simply ignore
the directions in our analysis and apply the eigenvector calculation to the
undirected network.  This approach has been taken before by other authors
and appears to work well---see, for example, Ref.~\cite{Krause03}.

\begin{figure}
\includegraphics[width=8cm]{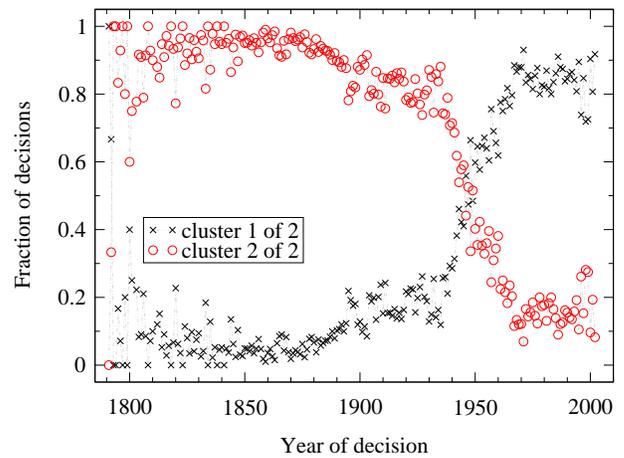}
\caption{A histogram of the number of decisions versus the year of the
  decision for cases assigned to each group in the two-way split produced
  by the modularity maximization algorithm.}
\label{fig:clusteringHistograms1}
\end{figure}

We can visualize the results of our clustering analysis in a manner similar
to our visualizations of the output of the EM algorithm, as a histogram
over time.  The results for the leading split of the Supreme Court network
into two clusters are depicted in this way in
Fig.~\ref{fig:clusteringHistograms1}.  The results are similar to those for
the EM algorithm, with a significant break around 1937.  This appears to
bolster the conclusions of our EM analysis, that there have been separate
periods in the court's history that leave identifiable signatures in the
citation record.  There are some differences between the two sets of
results, particularly the early ``tail'' to the second group in the
clustering analysis and an overall difference in the number of cases
assigned to each group.  A possible explanation for these differences is
that the EM analysis makes use only of citations received by cases, whereas
the clustering analysis, which ignores edge direction, takes into account
both citations received and citations made.  This allows the classification
into groups of some vertices that were unclassifiable with the EM algorithm
by virtue of never receiving any citations.  (About 10\% of cases were
never cited.)  It could also be responsible for the tail in the second
group because citations made, which are necessarily to cases in the past,
connect vertices to earlier times, perhaps pulling them from the second
group into the first in the clustering analysis.

\begin{figure}
\includegraphics[width=8cm]{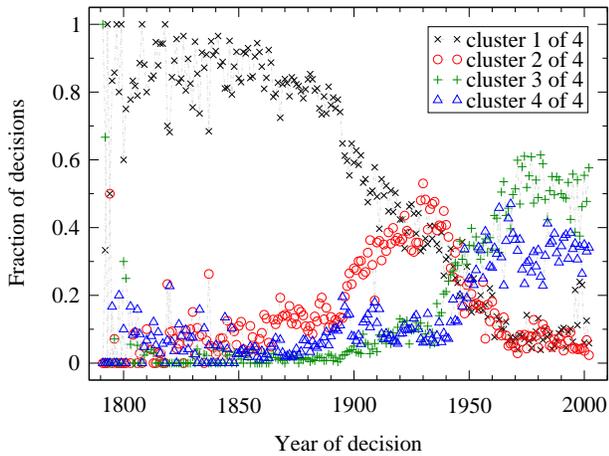}
  \caption{A histogram of the number of decisions versus the year of the
    decision for cases assigned to each group in the four-way split
    produced by the modularity maximization algorithm.}
\label{fig:clusteringHistograms2}
\end{figure}

As with the EM analysis, we can go further and look at splits into larger
numbers of groups.  For instance, Fig.~\ref{fig:clusteringHistograms2}
shows the best split into four groups according to the modularity-based
approach.  Again the split is similar in overall form to the split found by
the EM algorithm with $c=4$, although the results are not as clean as those
for the EM algorithm.  As before, a new split point appears around 1900,
which could be associated with the start of the \textit{Lochner} era.

\section{Vertex authority score and time evolution}
\label{sec:authority}

For our third analysis, we turn away from studies of groups or clusters and
focus on another class of network measures: centrality scores, which
quantify the importance or influence of individual vertices in a network.
As we will see, the pattern of centrality scores as a function of time in
our evolving citation networks can reveal interesting patterns.

The simplest of centrality scores is the degree of a vertex.  In a directed
network such as a citation network, there are two degrees, the in-degree
and the out-degree.  It is reasonable, for instance, to imagine that
important or influential vertices in a citation network will receive many
citations and therefore have high in-degree.  A more sophisticated versions
of the same idea is eigenvector centrality~\cite{Bonacich87}, in which,
rather than merely counting the number of citations a vertex gets, we award
a higher score when the citing vertices are themselves influential.  The
simplest way to do this is to define the centrality to be proportional to
the sum of the centralities of the citing vertices, which makes the
centralities proportional to the elements of the leading eigenvector of the
adjacency matrix.  Unfortunately, this method does not work for acyclic
directed networks, such as citation networks, for which all such
centralities turn out to be zero.

An interesting variant of eigenvector centrality has been proposed by
Kleinberg~\cite{Kleinberg99a} that works well for acyclic networks.  In
this variant each vertex has two centralities, known as the \defn{authority
  score} and the \defn{hub score}, the first derived from the incoming
links and the second from the outgoing links.  In this view a ``hub'' is a
vertex that points to many important authorities---a review paper in a
citation network, for instance---while an authority is a vertex pointed to
by many important hubs---such as an important or authoritative research
article on a particular subject.  In the simplest version of the method the
authority score~$x_i$ of vertex~$i$ is simply proportional to the sum of
the hub scores~$y_j$ of the vertices citing it:
\begin{equation}
x_i = {1\over\lambda} \sum_j A_{ij} y_j,
\end{equation}
for some constant~$\lambda$, while the hub score is proportional to the sum
of the authority scores of the vertices it cites:
\begin{equation}
y_i = {1\over\mu} \sum_j A_{ji} x_j.
\end{equation}
In matrix form, these equations can be written
\begin{equation}
\mathbf{Ay} = \lambda \mathbf{x},\qquad
\mathbf{A}^{\!T}\mathbf{x} = \mu \mathbf{y}.
\end{equation}
Or, eliminating either $\mathbf{x}$ or $\mathbf{y}$,
\begin{align}
\label{eq:authorities}
\mathbf{AA}^{\!T}\mathbf{x} &= \lambda \mu \mathbf{x},\\
\label{eq:hubs}
\mathbf{A}^{\!T}\mathbf{Ay} &= \lambda \mu \mathbf{y}.
\end{align}
Thus $\mathbf{x}$ and $\mathbf{y}$ are eigenvectors of the symmetric
matrices $\mathbf{AA}^{\!T}$ and $\mathbf{A}^{\!T}\mathbf{A}$ (also known
as the \defn{cocitation} and \defn{bibliographic coupling} matrices
respectively).  In Kleinberg's formulation of the problem one focuses on
the leading eigenvector of each of the matrices, although in principle
there could be useful information to be gleaned from other eigenvectors
too.

\begin{figure}
\includegraphics[width=8cm]{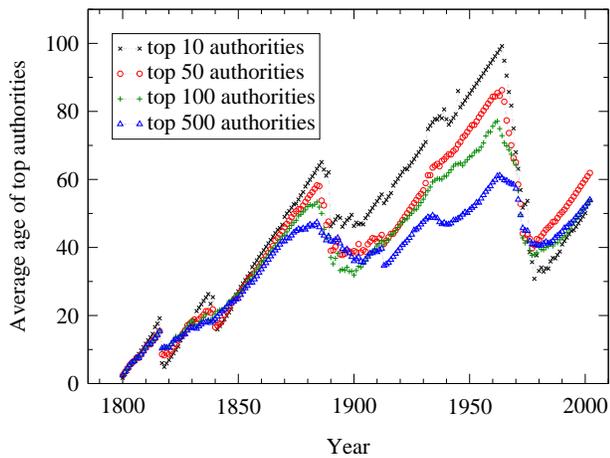}
\caption{The average age of the highest-authority cases in the Supreme
  Court citation network as a function of time.}
\label{fig:authorityScores}
\end{figure}

Taking the Supreme Court network as an example again, we have applied this
method to the calculation of authority scores for cases in the network.  It
proves particularly revealing to look at the scores as a function of time.
That is, we take the network as it existed at some time~$t$ (discarding all
cases published after that time) and calculate a complete set of authority
scores for all vertices.  We concern ourselves primarily with the most
central cases, those with the highest scores.
Figure~\ref{fig:authorityScores} shows one particularly revealing
statistic, the average age of the ten highest-ranked cases for each year in
our data set as a function of year.  As the plot shows, there is a marked
trend for the average age to increase in step with the passage of time.
This is precisely the behavior one would expect if the top authorities in
the network are remaining the same as time goes by.  Every once in a while,
however, the plot shows a sudden and precipitous drop in the average age,
indicating that a much younger set of vertices have, in a short space of
time, taken over as the new leaders in the authority score rankings.  Thus
the plot indicates a repeated pattern in the evolution of the network in
which a certain set of vertices---certain cases considered by the Supreme
Court---remain the top authorities for substantial periods of time before
being swiftly replaced by a different set.  One example of such a turnover
can be seen in Fig.~\ref{fig:authorityScores} around 1900 and a smaller one
around 1940, dates that, as we have seen, correspond roughly to the
beginning and end of the \textit{Lochner} era.  Another very large dip in
the curve occurs around 1970.  (Our four-group EM analysis also found a
group division at approximately the same time---see
Fig.~\ref{fig:emsplit3}.)  The large size of this dip may be due in part to
the much larger number of cases decided per year by the Supreme Court in
more recent decades than in its earlier history, which makes it easier for
newly appearing cases to quickly become top authorities.  The results of
the centrality analysis are thus compatible with but different from those
of previous sections.  Such variations are one reason why a variety of
different analytic techniques are useful in studies of network structure.

The behavior described is clearest in the age of the top ten vertices, but
persists if a different number is used.  Figure~\ref{fig:authorityScores}
shows the results of the same calculation for the top 50, 100, and 500
authorities, and in each case a similar pattern of maturation followed by
swift renewal is visible.

\section{Discussion}
\label{sec:discussion}
Although the purpose of this paper is primarily to highlight new methods
for the analysis of network data, the ultimate goal of these methods is of
course to give researchers insight into the structure and meaning of their
data.  Thus it is interesting to ask whether the analyses described here do
in fact shed light on the system studied---in this case, the network of
citations between Supreme Court cases.  In fact the results do appear to
shed interesting new light on the workings of the Supreme Court; we give a
short explanation of our arguments in this section.

The United States underwent a transition from an agricultural economy to an
industrial economy in the latter part of the nineteenth century.  Federal
and state legislators adapted to the new economic environment by passing
laws that regulated emerging industries.  These regulations, however, were
not without opposition from those who preferred a \textit{laissez-faire} or
hands-off approach.  Among those outspoken in opposition were several
members of the Supreme Court and, beginning in 1897, the court began
invalidating a number of cases that imposed regulations on industry and
business, starting with \textit{Allgeyer v.\ Louisiana}.  The legal
doctrines of \defn{substantive due process} and \defn{freedom of contract}
were merged together into a significant limitation on the police power of
the state.  After \textit{Allgeyer}, any statute, ordinance, or
administrative act that imposed any kind of limitation upon the right of
private property or freedom of contract became suspect, even if the
regulation was intended to promote safety and general welfare~\cite{KHB91}.

The most famous (or infamous) of the cases to use substantive due process
to invalidate state regulation was \textit{Lochner v.\ New York} in 1905, a
case that became so notorious that this entire era of jurisprudence,
between 1897 and 1937, came to be known as the \textit{Lochner} era.
During the \textit{Lochner} era the Supreme Court struck down nearly 200
regulations~\cite{UF02}.  The \textit{Lochner} era is clearly visible, for
example, in our EM analysis with $c=4$ (Fig.~\ref{fig:emsplit3})---the
analysis picks out one group of cases with start and end dates that
correspond closely to the accepted dates of the era.

Ultimately, the Supreme Court's hostility to state and federal regulation
began to interfere with the ``New Deal'' programs instituted by US
President Franklin Roosevelt to combat the Great Depression.  Between 1934
and 1936, the court invalidated more federal statutes than during any other
two-year period in its history and by 1936 nearly all of the statutes
passed as part of the New Deal had been struck down.  In response,
Roosevelt launched in early 1937 a counter\-offensive against the Supreme
Court in which he proposed to appoint to the court up to six additional
justices more receptive to the New Deal.  This ``court packing'' plan was,
to say the least, highly controversial, but Roosevelt had the support of
significant majorities in both houses of Congress, and the nation as a
whole, still in the throes of the depression, was eager for something new.

Following Roosevelt's proposal, the court abruptly reversed course and,
beginning in March of 1937, validated a series of state and federal
measures.  Contemporary commentators have humorously dubbed this change the
``switch in time that saved nine,'' but whether the switch was substantive
or illusory has been the subject of much debate.  Some scholars believe
that the court responded to political pressure, while others have suggested
that the court already contained a majority of justices who would have been
inclined to sustain the New Deal if legislation had been drafted better or
if certain unanswered questions had been appropriately posed to the court.

Our EM analysis shows a clear break around 1937, corresponding closely to
the end of the \textit{Lochner} era.  It is important to appreciate that the
analysis takes into account only citations received by cases, and thus that
the opinions of the Supreme Court appear to have taken a substantial change
of direction not merely in impact but also in their arguments: later cases
cited the new opinions rather than those coming before them because,
presumably, their arguments better supported the decisions of the post-1937
court.  Thus our analysis appears to indicate not merely a change in case
outcomes that was a natural, if novel, result of positions long held by the
sitting justices, but a more fundamental change in legal thinking
itself---or at least its expression in the written opinions of the court
and the later citation of those opinions.

\section{Conclusions}
In this paper we have described several methods for the analysis of
citation networks, which are acyclic directed graphs of citations between
documents.  Using the network of citations between opinions handed down by
the US Supreme Court as an example, we have described and demonstrated
three analysis techniques.  The first makes use of a probabilistic mixture
model fitted to the observed network structure using an
expectation--maximization algorithm.  The second is a network clustering
method making use of the recently introduced method of modularity
maximization.  The third is an analysis of the patterns of time variation in
eigenvector centrality scores, particularly the ``authority'' score
introduced by Kleinberg~\cite{Kleinberg99a}.

When applied to the Supreme Court network, each of these analyses reveals
interesting structure, particularly highlighting qualitative changes in
citation patterns that may be associated with specific eras of legal
thought in the Supreme Court.  However, it is in combination that the
methods become most effective.  Features that appear clearly in analyses
performed using several different techniques possess correspondingly
greater persuasive force.  In the case of the Supreme Court, there emerges
quite a clear picture of the eras of the court as marked by shifts in
citation patterns, particularly around the time of the so-called
\textit{Lochner} era in the early 20th century.

\begin{acknowledgments}
  This work was funded in part by the National Science Foundation under
  grant number DMS--0405348 and by the James S. McDonnell Foundation.
\end{acknowledgments}

\end{document}